# Energy-Efficient FPGA Framework for Non-Quantized Convolutional Neural Networks

Angelos Athanasiadis[1], Nikolaos Tampouratzis[2], Ioannis Papaefstathiou[1]

Department of Electrical & Computer Engineering, Aristotle University of Thessaloniki, Greece[1]
Department of Industrial Engineering & Management, International Hellenic University, Greece[2]
Email: angelathan@ece.auth.gr; ntampouratzis@ihu.gr; ygp@ece.auth.gr

*Abstract*—The growing demand for real-time processing in artificial intelligence applications, particularly those involving Convolutional Neural Networks (CNNs), has highlighted the need for efficient computational solutions. Conventional processors, very often, fall short in balancing performance, power consumption, and latency, especially in embedded systems and edge computing platforms. Field-Programmable Gate Arrays (FPGAs) offer a promising alternative, combining high performance with energy efficiency and reconfigurability. The presented framework addresses the complex and demanding computations of CNNs on FPGAs maintaining full precision in all neural network parameters. Specifically, our framework is based on Darknet which is very widely used for the design of CNNs and allows the designer, by using a similar input to that given to Darknet, to efficiently implement a CNN in a heterogeneous system comprising of CPUs and FPGAs. When compared with the FPGA frameworks that support quantization, our solution aims to offer similar performance and/or energy efficiency without any degradation on the NN accuracy.

## I. INTRODUCTION

In the recent years, Convolutional Neural Networks (CNNs) are the key components for many significant advancements in the field of artificial intelligence. They have proven to be highly effective in numerous fields like image, video and natural language processing. Additionally, they are effective in various tasks including image classification, object detection, and semantic segmentation. It often encounters difficulties like high processing power and power consumption. The complicated structures of CNNs, which consist of several convolutional layers, fully connected layers, and a large number of parameters, require significant computational resources which can be a substantial barrier, particularly for applications that demand real-time processing and need to be deployed on devices with limited resources, such as embedded systems and edge computing platforms.

Various approaches [1],[2],[3] have been proposed to implement CNNs focusing mainly in heavily quantized models. Non-quantized models maintain the full precision of the network parameters, ensuring high accuracy at the cost of higher resource utilization and power consumption. Quantized models, on the other hand, reduce the precision of the parameters, thereby lowering resource usage and power consumption but very often at the cost of reduced accuracy.

In this paper we present a novel design and implementation framework that allows for the seamless FPGA implementations of non-quantized CNNs with high performance, energy efficiency and accuracy. The main benefits of our approach are:

- **Accuracy Preservation:** By avoiding quantization and retaining full precision, the proposed framework aims to preserve the accuracy of the CNN models.
- **High Design Productivity, Flexibility and Adaptability:** The presented efficient design flow is based on the widely used DarkNet CNN design framework, it is based on purely C/C++ and targets the whole range of FPGAs from the smallest to the largest ones.
- **High Performance:** The proposed framework can fully exploit the parallelism of any FPGA to accelerate the inference process of CNNs, ensuring timely and efficient processing.
- **Energy Efficiency:** The proposed framework optimizes the power efficiency of CNN inference on FPGAs, making it suitable for power-sensitive applications.

## II. FRAMEWORK ARCHITECTURE

Figure 1 presents the framework architecture of our framework which handles convolutional and deconvolutional layers within neural networks in an automated manner. This architecture is represented as a flowchart, emphasizing the sequential and parallel processes involved in the tool's operation.

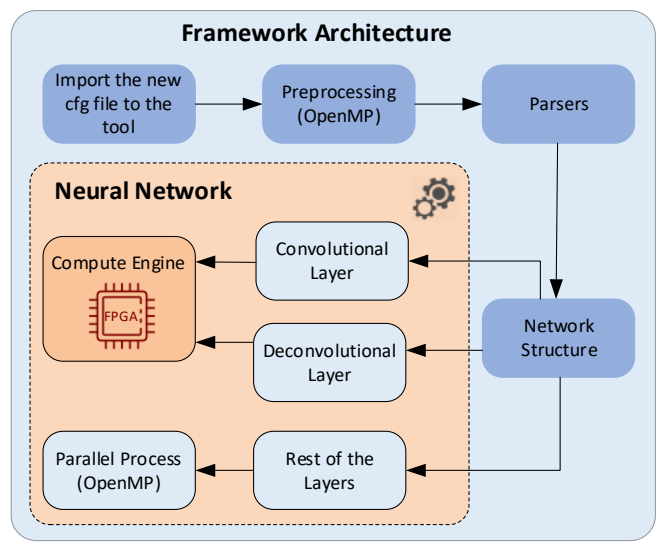

*Figure 1: Architecture of the Framework*

The key element is the Innovative Compute Engine using High Level Synthesis (HLS) on FPGA, which allows the execution of multiple mathematical executions in a single clock cycle achieving a relatively high throughput and performance.

## III. IMPLEMENTATION OF INNOVATIVE HLS COMPUTE ENGINE

Our HLS FPGA Kernel handles mainly the matrix multiplication tasks which is the cornerstone in virtually every CNN implementation. Moreover, as illustrated in Figure 2, we utilize the internal BRAMs in conjunction with HLS streams, to optimize also the on-chip memory access patterns and further increase the computational efficiency.

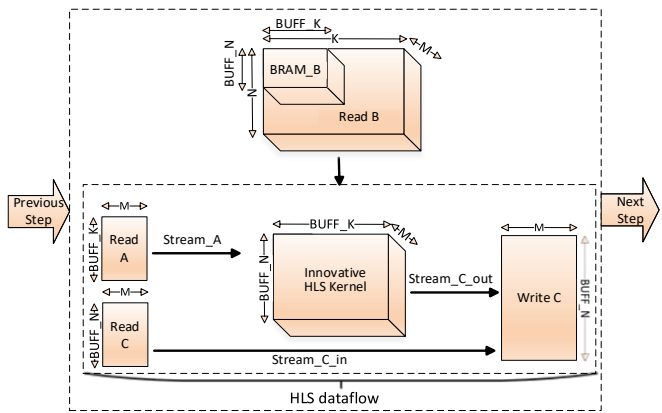

*Figure 2: Architecture of Innovative Compute Engine*

The aforementioned outcome leads to a significant decrease in the effective latency associated with off-chip memory retrieval and guarantees the accessibility of data for computational purposes, thereby diminishing idle time caused by external memory latency. Specifically, streams enable a continuous flow of data between processing elements without the need for intermediary storage in BRAMs, minimizing the latency and resource overhead. By leveraging streams, data is transferred directly between producer and consumer processes, facilitating pipelined execution and enhancing parallelism. This direct transfer mechanism reduces the dependency on BRAMs for temporary storage, leading to more efficient resource utilization and higher throughput. Furthermore, streams can handle variable data rates more effectively ensuring that processes are not overwhelmed by the rate of data production.

In addition, modern FPGAs are connected to multiple memory banks with dedicated channels (e.g., multiple DRAM modules or HBM lanes) in order to increase the external memory bandwidth. In order to take advantage of those multiple lanes/modules we insert the appropriate HLS directives that split data transfers into a parameterizable number of memory banks/lanes so as to take full advantage of the bandwidth available in each of the memory interfaces. The transfers of data between the central processing unit (CPU) and the field-programmable gate array (FPGA) are facilitated at the full data width (512 bits) per clock cycle, thereby ensuring high-throughput communication between the processing elements and memory subsystems.

## IV. EXPERIMENTAL RESULTS

To demonstrate further the effectiveness of the presented approach we compare the non-optimized reference model, the fully optimized model on both a high-end FPGA (AMD Alveo U55C) and an embedded one (Kria KR260), the parallel execution of matrix multiplication using OpenMP in a multicore CPU and using CUDA on an NVIDIA T4 GPU. In all experiments the array dimensions which are M=2048, K=4096, and N=16384 are selected so as to be different from those triggering our peak performance demonstrating the flexibility of our approach (since it can effectively handle any shape of matrices).

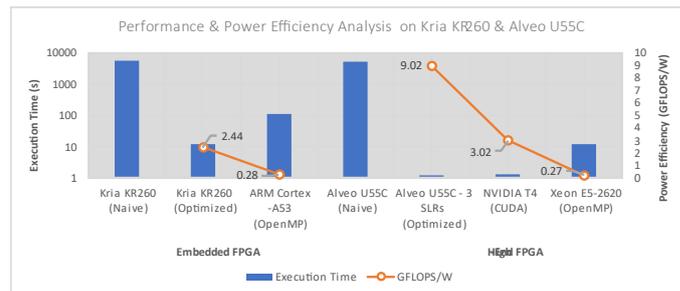

*Figure 3: Performance and Power Efficiency Analysis (FP32)*

Furthermore, the results obtained from numerous experiments with different dimensions are fully inline with those presented in Figure 3. As illustrated in this figure, our fully-optimized implementation for the embedded FPGA is two orders of magnitude faster than the reference implementation and 9x times faster compared to the fully parallelized algorithm executed on an embedded ARM 4core CPU (Cortex-A53); those numbers include all the memory accesses and the external memory technologies and topologies are exactly the same. Similarly, our fully optimized approach when implemented on the Alveo U55C board is approximately three orders of magnitude faster than the reference implementation and 10x times faster compared to the fully parallelized algorithm executed on an Intel Xeon E5-2620 v4 (8 cores). In order to further compare the overall efficiency of the presented approach with that triggered by the software implementation, the GFLOPS/Watt for each implementation were also measured. Based on our measurements we achieve 34x and 9x higher energy efficiency than the best CPU parallel implementation in an Alveo U55C and a Kria KR260 respectively. Moreover our design, when implemented on the Alveo, is 3x more power efficient than the CUDA implementation on an NVIDIA T4 GPU which is also implemented on a better CMOS technology (12nm for T4 vs 16nm for U55c).

## V. CONCLUSION

In conclusion, this research addresses the critical need for efficient CNN implementations in power-constrained environments. The proposed non-quantized FPGA-enabled CNN framework successfully combines high performance with energy efficiency, leveraging the inherent parallelism and reconfigurability of FPGAs. By maintaining full precision in network parameters, the framework achieves high accuracy without compromising on resource utilization or power consumption. The experimental results validate the framework's effectiveness, showing substantial speedups in inference processing and significant reductions in power usage.